\documentclass[final]{IEEEtran}
\usepackage{amsthm,amssymb,graphicx,multirow,amsmath,color,amsfonts}%,ulem}
\usepackage[update,prepend]{epstopdf}
\usepackage[noadjust]{cite}
\usepackage[latin1, utf8]{inputenc}
\usepackage{tikz}
\usepackage{bbm} % for \mathbbm{1}
\usepackage{pdfpages}
\usepackage{tabulary}
\usepackage{multirow}
\usepackage{comment}
\usepackage{subfigure}
\usepackage{float}

\def\nb0{{\mathbf{0}}}
\def\nb1{{\mathbf{1}}}

\allowdisplaybreaks % Allows breaking of eqnarray over multiple pages (avoids unnecessary blanks in the document before eqnarray)
\begin{document}
	\title{
		Drone Charging Stations Deployment in Rural Areas for Better Wireless Coverage: Challenges and Solutions
	}
	\author{
		Yujie Qin, Mustafa A. Kishk, {\em Member, IEEE}, and Mohamed-Slim Alouini, {\em Fellow, IEEE}
			\thanks{Yujie Qin, Mustafa A. Kishk, and Mohamed-Slim Alouini are with Computer, Electrical and Mathematical Sciences and Engineering (CEMSE) Division, King Abdullah University of Science and Technology (KAUST), Thuwal, 23955-6900, Saudi Arabia (e-mail: yujie.qin@kaust.edu.sa; mustafa.kishk@kaust.edu.sa; slim.alouini@kaust.edu.sa).} 
	}
	
	\maketitle
	\begin{abstract}
		While the fifth-generation cellular (5G) is meant to deliver Gigabit peak data speeds, low latency, and connect to billions of devices, and 6G is already on the way, half of the world population living in rural areas are still facing challenges connecting to the internet. Compared with urban areas, users in rural areas are greatly impacted by low income, high cost of backhaul connectivity, limited resources, extreme weather, and natural geographical limitations. Hereby, how to connect the rural areas and what are the difficulties of providing connectivity draw great attention. % 
		This article first provides a brief discussion about existing technologies and strategies for enhancing the network coverage in rural areas, their advantages, limitations, and cost. Next, we mainly focus on the UAV-assisted network in resource-limited regions. Considering the limitation of the on-board battery of UAVs and the electricity supply scarcity in some rural regions, we investigate the possibility and performance enhancement of the deployment of renewable energy (RE) charging stations. We outline three practical scenarios, and use simulation results to demonstrate that RE charging stations can be a possible solution to address the limited on-board battery of UAVs in rural areas, specially when  they can harvest and store enough energy. Finally, future works and challenges are discussed.
	\end{abstract}
	%-------------------------------------------
	\section{Introduction}
%digital divide
The issue of 'digital divide', which refers to the technological gap between those who benefit from the digital age, revolutionary changes and those who do not, continues to widen \cite{van2006digital}. Rural residents are at high risk of being marginalized in the information-based economy and advanced technological innovations, as a result of the uneven development between urban and rural areas. A large section of rural residents is on the wrong side of the digital divide.
For instance, information and communication technology (ICT) is expected as one of the future's basic technologies \cite{8570906}. The internet of things (IoT), considered as the backbone of the network of smart cities, tends to integrate people, devices, sensors and systems to achieve smart grids, energy-efficient environments and intelligent transportation systems \cite{9301389}. All these technologies rely on the deployment of  efficient communication networks, however, rural residents are more likely to suffering from the connection issues or even unable to have internet access, and wired networks are usually too expensive to reach some remote areas \cite{9301389,9042251}. The low incomes of residents and absence of the basic infrastructure in resource-limited rural areas undermines the diffusion of ICTs  and the development of IoT. 

%diff rural and urban
\subsection{Limited Resources}
Rural areas tend to differ from urban areas in terms of economic attainments and the available physical infrastructure. Compared with urban communications, rural networks are highly limited by constrained resources, such as the lack of reliable electricity supply. Generally, electricity is provided by the grid. However, the grid infrastructure is often poor due to the financial budget, extreme weather conditions, and geographic nature in remote areas, such as large scales of the desert or forest, and high cost, including the high maintenance and operation cost, and transportation and fuel storage expenses. More problematically, the aforementioned reasons increase the difficulty of obtaining a return on investment (ROI) of the service provider, resulting in a vicious circle that worsens the condition.
To compensate the absence of physical infrastructure, applying alternative renewable energy sources, such as solar and wind, is one of the possible solutions. Solar power is currently widely used to power household electrical needs. For example, the Renewable Energy Development Project (REDP), launched by Chinese Government used solar and wind technologies to supply the electricity in remote areas and institutions, which covers millions of household systems. As a result, implementing the solar system, wind generators and combining them with storage capacities are eco-friendly and economically advantageous compared with high-cost and grid-based electricity. 

\subsection{Unique User Distribution}
Apart from less available infrastructure and limited power supply, the locations of users exhibit unique properties. Typically, rural areas have a low density of population and medium size settlements, as well as a large area of agriculture. On the one hand, rural residents are more clustered, since a group of people live and work together. On the other hand, they are more scattered, due to the large scale of working areas. For instance, while rural residents live together in one village, their farms may locate scattered around the village and very far from each other. Hence, the spatial distribution of rural users is non-uniformly distributed and highly dynamic, e.g., higher in the cluster center and lower at edge at night and opposite while working. This, in turn, increases the probability of outage of edge users when a large number of residents are working in farms that are far from the cluster center, and the cluster base station (BS) can only cover a certain area. 
In addition, the locations of these medium-size settlements also exhibit a similar property of clustering: the medium-size settlements are scattered around the large-size user clusters, such as some villages scattered around a town.  Large-scale distribution and dynamic movements of rural users increase the difficulties of delivering cellular network coverage greatly.

In this article, we investigate the possibility of combing renewable energy in rural areas to address the UAV battery	lifetime issue, and explore three practical scenarios which capture the potential challenges of this system. The reminder of this article is organized as follows. Current challenges and approaches are described in Section II. Detailed system models and simulation results are discussed in Section III. Future works and open problem are included in Section IV.
%Hence, we adapt the Thomas Cluster Process (TCP) model in the above three scenarios, in which users are distributed according to a symmetric distribution with variance $\sigma_{u}^2$.

%in rural area tend to be more clustered with lower density and larger radius. 

%Drones
\section{Coverage Enhancement in Rural Areas}
Based on the above discussion, choosing scalable, cost-effective, and fast-deployment communication technologies is of vital importance. Multiple solutions have been provided in the literature to enhance the network cellular coverage in rural areas. In the following part of this section, we provide a brief discussion about these solutions based on different types of strategies, and some interesting projects. 

{\em Satellites.}
Satellite backhaul is considered to be the basic technology that connects rural areas. It overcomes the challenge of geographic constraints, weather conditions, and the most important, the absence of available infrastructures (no need for towers and grids). It can provide direct communication links with users in remote areas. However, the high cost of leasing bandwidth and relatively high latency, compared with other technologies, should be considered. Recently, some researchers are working on the reduction of the latency and investigating the possibility of using  Transmission Control Protocol acceleration. Besides, the bandwidth cost is expected to decrease with the increasing deployment of High Throughput Satellites (HTSs), please refer to \cite{9042251} for more details.

{\em  Optical Fiber Link.} 
Optical fiber link is based on cables and typically ensures long-distance communications and the highest bandwidth. Compared to electricity in electrical cables, the attenuation of light propagates through the fiber is much lower, which allows a large-scale deployment. However, in the case of sparse rural areas, 
the maintenance cost will be relatively high, and ROI is very low. This is because rural areas often face more extreme weathers, natural obstacles, sparse population, fewer subscribers, and consequently, higher cost and lower return. To minimize the cost, several optimization approaches are provided, such as deploying fibers along the railroad or power lines to reduce the cost in infrastructure, and finding the minimum spanning tree of point to multi-point communications \cite{dalela2014geo}, hence minimizing the length of fibers.

{\em Microwave and Free Space Optics (FSO).}
When fiber deployment is too expensive and difficult to reach some remote areas, wireless solutions can be a possible solution, such as microwave and FSO. Microwave link is based on radio frequency (RF) equipment which is deployed on the top of buildings or high towers. To reach the remote rural areas, however, appropriate towers carrying the equipment need to be built, which highly increases the operation expenditure cost. In practical scenarios, microwave towers' separation distances and coverage area can be properly designed to minimize the cost (e.g., based on weather, geographic conditions, and the area of interest). Besides, microwave link requires line-of-sight (LoS), which reflects the importance of the environment and weather conditions.

Generally, FSO includes terrestrial FSO and vertical FSO. Compared with vertical FSO, terrestrial FSO is more similar to microwave links which are based on deploying towers. However, FSO equipment is more expensive, and separation distance is relatively shorter. In the case of vertical FSO, it can communicate with the satellite and high altitude platforms, such as balloons, airships, and UAVs. Since high altitude platforms (HAPs) hover at a high altitude, they are less sensitive to weather conditions, and consequently, the separation distance of the vertical FSO can be larger. Authors in \cite{9301389} proposed intelligent radio resources management and network planning techniques and investigated the cost-efficiency of these strategies. Based on their results, the method based on vertical FSO with solar-powered devices is a suitable solution.

{\em TV band White Space (TVWS).}
TV white space is an interesting technology that has a good potential to enhance the coverage in rural areas. In the past, white spaces refer to the unused radio frequencies between active ones and are mainly used as buffers to protect broadcasting interference. In 2010, FCC approved the unlicensed public use of this spectrum, and now it is expected to provide broadband internet access with surrounding TV channels. In comparison with WiFi, TVWS can cover around 10 kilometers in diameter, which is 100 times larger, and the uneven ground, owing to the ability of penetrating the obstacles, such as trees, and no requirement of LoS channels. This technology works well in rural areas because of the fewer people and less saturated airwaves, and it can be used to connect sensors, IoT devices, and ensure point-to-point connectivity in the field. While TVWS is a potential option to enhance the coverage and transfer data in rural areas, available infrastructure, intergovernmental operations and public-private partnership are still challenges.

{\em HAPs.}
A large amount of existing works investigate using HAPs to connect the rural areas. The most popular types of HAPs are balloons, airships, and gliders. In addition to being less sensitive to the weather conditions, the main advantages of these platforms are wide coverage areas, ease of deployment and can be used in disaster environment. Since they are deployed very high (about 22 km), hovering energy is actually greatly reduced due to the low wind currents and turbulence. Moreover, such equipment can be fully charged by solar power because of exposure to the sun. Therefore,the operational cost can be greatly reduced. Many recent projects over the globe focus on using HAPs for coverage enhancement such as Google Loon project \cite{burr2017feasibility }. It aimed to connect rural areas by launching the high-altitude super-pressured solar balloon in the stratosphere. However, implementing such technology is still too expensive.
Currently, high cost and the complexity of the devices are still the main limitations.

{\em Unmanned Aerial Vehicles (UAVs).} 
The UAV-assisted communication system attracts great attention among all the current works, owing to its unique attributes, such as mobility and easy deployment.  
In comparison with the existing terrestrial base stations (TBSs), UAVs can efficiently function as aerial BSs (ABSs) with high relocation flexibility based on the dynamic demand of rural users \cite{qin2021influence}. They can optimize their locations by tracking the dynamic movements of rural users, then maximize the coverage probability.  In addition, compared with urban areas, there are fewer high buildings and other obstacles. Hence, deploying UAVs in rural areas is more likely to yield better communication channels due to the high probability of establishing  LoS   links and lower additional loss  with ground users \cite{al2014optimal}.

%agriculture and covidrone
Besides communication, UAVs are widely applied in agriculture as remote sensing in rural areas. Some primary functions include collecting high-quality images about the fields and animals and using these photos to investigate the vegetation status, such as mapping inland water, loss of vegetation and moisture content in the plant, and wildlife habits or damage. 

More attractively, a current study proposes a drone-based contact-less COVID-19 diagnosis and testing in rural areas named Covidrone \cite{chintanpalliiomt}. In their system, the COVID-19 text kit is given to the patient having a high likelihood of infecting based on raw data collected by UAVs and analyzed by a deep neural network. Such work provides some interesting thoughts of combating COVID-19 in medical facilities limited rural areas.

%Battery
Despite the variety of benefits of UAVs, the limited on-board battery life is one of the primary system's bottlenecks. UAVs need to fly back to a nearby charging station to recharge/swap their battery frequently. Hence, the service UAV offers is likely to be interrupted once their energy runs low, and users in UAVs' coverage areas experience poor service quality. This, in turn, restricts the performance of the UAV-enabled wireless network.  Authors in \cite{8761318} proposed the use of solar-powered charging stations to address the energy limitation of UAVs. In \cite{9446593}, Lyapunov optimization is used to schedule the charging and energy management of UAVs and charging stations. In our previous work  \cite{qin2021influence}, we study the influence of charging stations on the system performance in urban areas. We first compute the impact of the UAVs' limited battery, the density of the charging stations and the recharging time on the performance of the considered setup, which do have a high impact on the coverage probability. The results revealed that we can achieve a similar coverage probability by deploying lower dense charging stations if we can reduce the charging time. Similarly, in urban areas, we then analyze the influence of charging stations' capacity. The network's performance degrades as the capacity of charging stations decreases or the number of UAVs served by one charging station increases. The results showed the importance of considering the traffic in charging stations and optimal scheduling, and a trade-off between deploying high density of charging stations with small capacity or deploying low density of charging stations with large capacity.

%we compute the influence of the spatial distribution of capacity-limited charging stations on the network performance in urban areas \cite{qin2021influence}. In this work, we considered the locations of UAVs and charging stations are two independent PPPs, and UAVs fly back to the nearest charging station. In the case of independent distribution, the association area of charging stations forms a Poisson Voronoi tessellation and the number of UAVs in the typical cell (the cell contains the origin) is a random variable. In this case, the availability of UAVs (the event that UAVs can provide service) depends not only on the remaining energy of battery, but also the traffic in the capacity-limited charging station (e.g., UAVs may need to wait in a queue to charge the battery and then fly back to provide service). The result shows a trade-off between a high dense of deploying and slightly increasing the density or the capacity of charging stations.

While we take the electricity limitation in rural areas into account, the deployment of the RE charging station becomes one of the possible solutions to the UAV-assisted network. This, thus, motivates this article to explore several scenarios where RE charging stations are deployed in rural areas. To do so, we simulate the system performance in three  scenarios including, with/without RE charging stations, limited capacity or low energy storage of RE charging stations, and a more realistic case that captures the dynamic movement of rural users. 

\section{Network Performance with Renewable Charging Stations}

Unless stated otherwise, we use the simulation parameters as listed herein Table \ref{par_val}. 
\begin{table}[ht]\caption{Table of Parameters}\label{par_val}
	\centering
	\begin{center}
		\resizebox{\columnwidth}{!}{
			\renewcommand{\arraystretch}{1}
			\begin{tabular}{ {c} | {c} | {c}  }
				\hline
				\hline
				\textbf{Parameter} & \textbf{Symbol} & \textbf{Simulation Value}  \\ \hline
				Density of medium-size user clusters & $\lambda_{\rm m}$ & $10^{-6}$ m$^{-2}$ \\ \hline
				Density of living and working user clusters & $\lambda_l$, $\lambda_w$ & $\lambda_w$ = $10\lambda_l$ \\\hline
				Battery size & $B_{\rm max}$ & $177.6$  W$\cdot$H \\\hline 
				Traveling-related power & $P_{\rm m}$ & 161.8 W \\\hline
				Service-related power & $P_{\rm s}$ & 177.5 W \\\hline
				UAV altitude & $h$ & 60 m\\\hline
				UAV velocity & $v$ & 10 m/s \\\hline
				Variance of user distribution & $\sigma_{u}^2$ & 120  \\\hline
				N/LoS environment variable & $a, b$ & 4.88, 0.43 \\\hline
				Transmission power of UAVs and TBSs& $\rho_{\rm u}$, $\rho_{\rm t}$ & 0.2 W, 10 W\\\hline
				SINR threshold & $\beta$ & 0 dB \\\hline
				Noise power & $\sigma^2 $ & $10^{-9}$ W\\\hline
				N/LoS, TBS path-loss exponent & $\alpha_{\rm n},\alpha_{\rm l},\alpha_{\rm t}$ & $4,2.1,4$ \\\hline
				N/LoS fading gain & $m_{\rm n},m_{\rm l}$ & $1,3$ \\\hline
				N/LoS additional loss& $\eta_{\rm n},\eta_{\rm l}$ & $-20,0$ dB 
				\\\hline\hline
		\end{tabular}}
	\end{center}
	%\vspace{-8mm}
\end{table}
As mentioned, rural users are more clustered in a large area and non-uniformly distributed compared with urban users. Therefore, among all the three scenarios we simulate in this article, the locations of rural users are modeled using a Thomas Cluster Process (TCP), in which users are distributed according to a symmetric distribution with variance $\sigma_{u}^2$, meaning that the locations of users are more dense in the center of clusters and perform like a 2D Gaussion distribution. Besides, we consider rotary-wing UAVs which can hover at a fixed altitude $h$ above the user cluster centers to provide service. Given the relative parameters in \cite{zeng2019energy}, we assume a fixed service-related power $P_{\rm s}$, which contains the hovering and total communication power, and traveling-related power $P_{\rm m}$. Assume that the reference user is located in a medium-size user cluster and associates with a UAV or a TBS which provides the strongest average received power. While environment variable $a,b$ of calculating the probability of LoS/NLoS model between the reference user and the serving UAV is given in \cite{al2014optimal}, the path-loss of both UAVs and TBS channels $\alpha_{\rm n},\alpha_{\rm l},\alpha_{\rm t}$, as well as other parameters and related communication models are given in \cite{qin2021influence}. 

\subsection{Deployment of Renewable Energy Charging Stations}

\begin{figure}[ht]
	\centering
	\includegraphics[width=1\columnwidth]{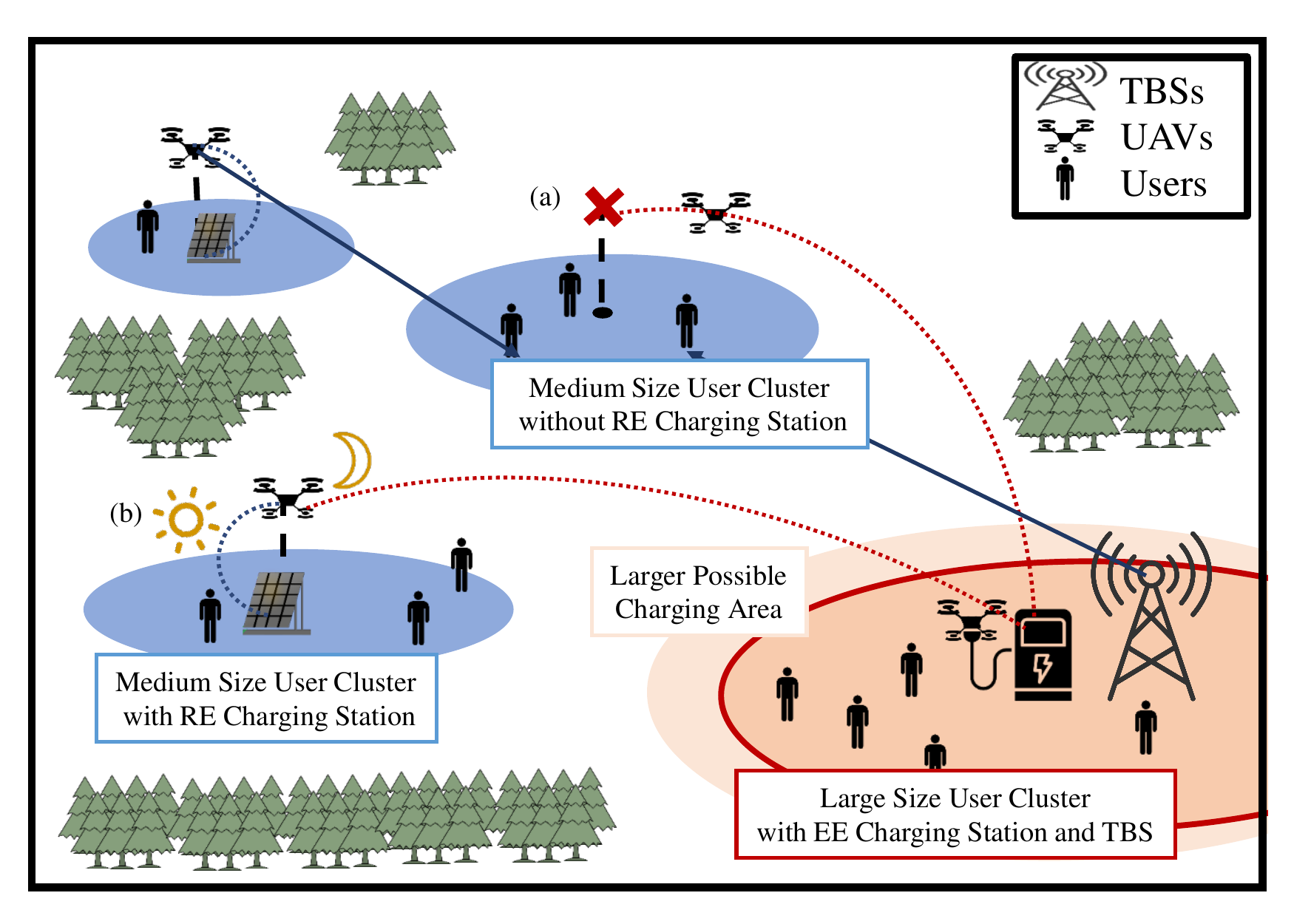}
	\caption{Illustration of the system setup.  (a) First scenario: UAVs need to fly back to EE charging stations if there is no RE charging station,  (b) Second scenario: UAVs can charge in RE charging stations during the day and may need to back to EE charging stations during the night.}
	\label{system_fig1}
\end{figure}

One of the most straightforward solutions to the limited battery lifetime is deploying more charging stations. However, given the electricity, infrastructure constrains and high cost, deploying a high dense of EE charging stations as urban areas becomes impossible. Consequently, deploying solar energy charging stations, as well as other renewable resources, becomes an alternative solution. In the first scenario, we consider some medium-size user clusters scattered around a large-size user cluster, as depicted Fig. \ref{system_fig1} (a). Assume that only the large-size user cluster has a TBS and electric energy (EE) charging station, and UAVs are deployed above the centers of medium-size user clusters to provide cellular network coverage. We compare the performances of whether the  RE charging stations are deployed at centers of user clusters or not. It means that when the battery of the cluster UAV gets drained, it can go to the RE charging station if there is one located at cluster center. If not, it needs to travel back to the EE charging station in the large-size user cluster. Besides, it may need to wait in a queue if there is more than one UAV in the EE charging station \cite{qin2021influence}.

%Due to the limitation of electric resources and infrastructures, we assume that only large-size user clusters have TBSs and electric charging stations, and UAVs are deployed above the centers of medium-size user clusters to provide cellular network coverage. Without loss of generality,  the capacity of all the charging stations in our system is finite and equals one. Therefore, when the battery of a cluster UAV gets drained, it can go to the RE charging station if there is one deployed in a medium-sized user cluster center. If not, it needs to travel back to the EE charging station in a large-size user cluster. Besides, it may need to wait in a queue if there is more than one UAV in the charging station.

%In this scenario, the number of the large-size user cluster is fixed at one and the locations of medium-size user clusters are modeled as a Poisson Point Process (PPP) $\Phi_m$, with density $\lambda_m$, the cluster UAVs are assumed to hover at a fixed altitude $h$ above each medium-size user cluster center, and the TBS, as well as all the charging stations, is located at the center of user clusters.

\begin{figure}[ht]
	\centering
	\includegraphics[width=1\columnwidth]{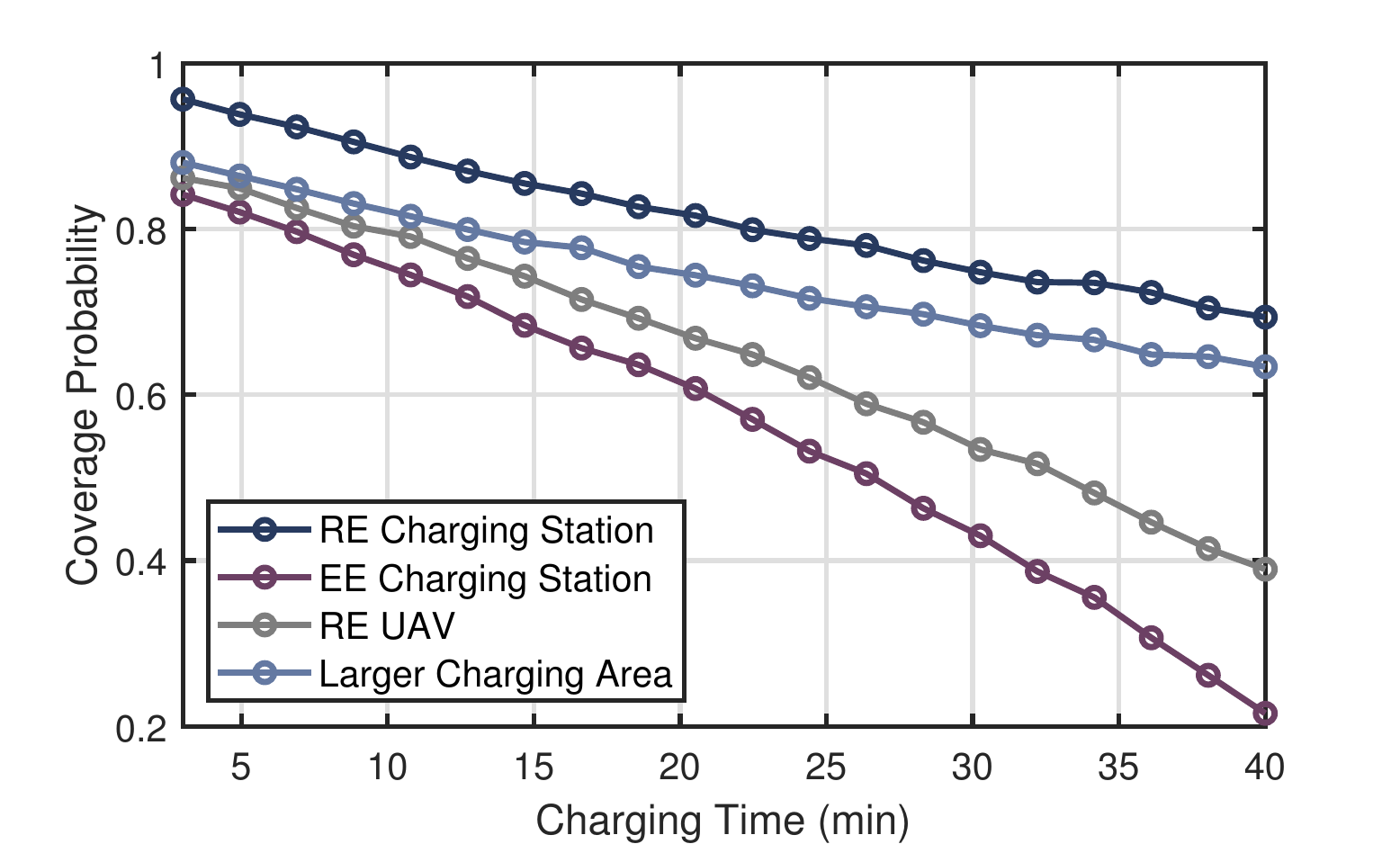}
	\caption{Coverage probabilities versus charging time under the different system deployment.}
	\label{fig_cov_dif_Tch}
\end{figure}

In Fig. \ref{fig_cov_dif_Tch}, we show the network performance under different charging station deployment strategies. As expected, the deployment of RE charging station at each medium-size user cluster center can enhance the network performance dramatically, especially when the charging time is long. Its performance is even better than one EE charging station for each cluster UAV, which is deployed at the edge of a larger possible area, e.g., assuming that the possible area of locating EE charging stations is constrained because of the limited grid infrastructure.
On the one hand, deploying RE charging stations highly reduces the traffic in the capacity-limited EE charging station, and decreases the traveling distance and energy consumption. Besides, we investigate the influence of using  RE UAVs, which can harvest a certain amount of energy from solar panel. However, the system performance does not improve a lot. The reason is that the power consumption model of UAVs is very sensitive to the total weight and rotor disc area. The solar panel equipped on UAVs cannot harvest too much energy but increases the total weight and the rotor disc area, which leads to more energy consumed during hovering and traveling.

\subsection{Limited Capacity of RE Charging Stations}
While the deployment of RE charging stations indeed shows a great improvement in the first scenario, the uncertainty is one of the main drawbacks of the RE system. Take solar system for example, the amount of harvested energy greatly depends on the weather and daytime. The output of a solar panel highly relies on sunlight radiation intensity, which is a function of day time, solar altitude angle and attenuation effect of clouds occlusion \cite{sekander2020statistical}. 
Once the sunlight is weak, the solar system cannot harvest enough energy, resulting a low energy stored in the battery. In this situation, the cluster UAV faces the problem of energy outage, and it always happens during the night-time.

Under the same system setup as the first scenario, we analyze the second scenario where the capacity or the harvested RE is limited, as shown in Fig. \ref{system_fig1}(b). This scenario captures the problem mentioned above that the cluster UAV needs to fly back to the EE charging station when the stored energy of the RE charging station runs low.

In the simulation, we use solar powered charging station as an example, but this result applies to all the RE sources: the stored energy in the RE charging station is limited. 

\begin{figure}[ht]
	\centering
	\includegraphics[width=1\columnwidth]{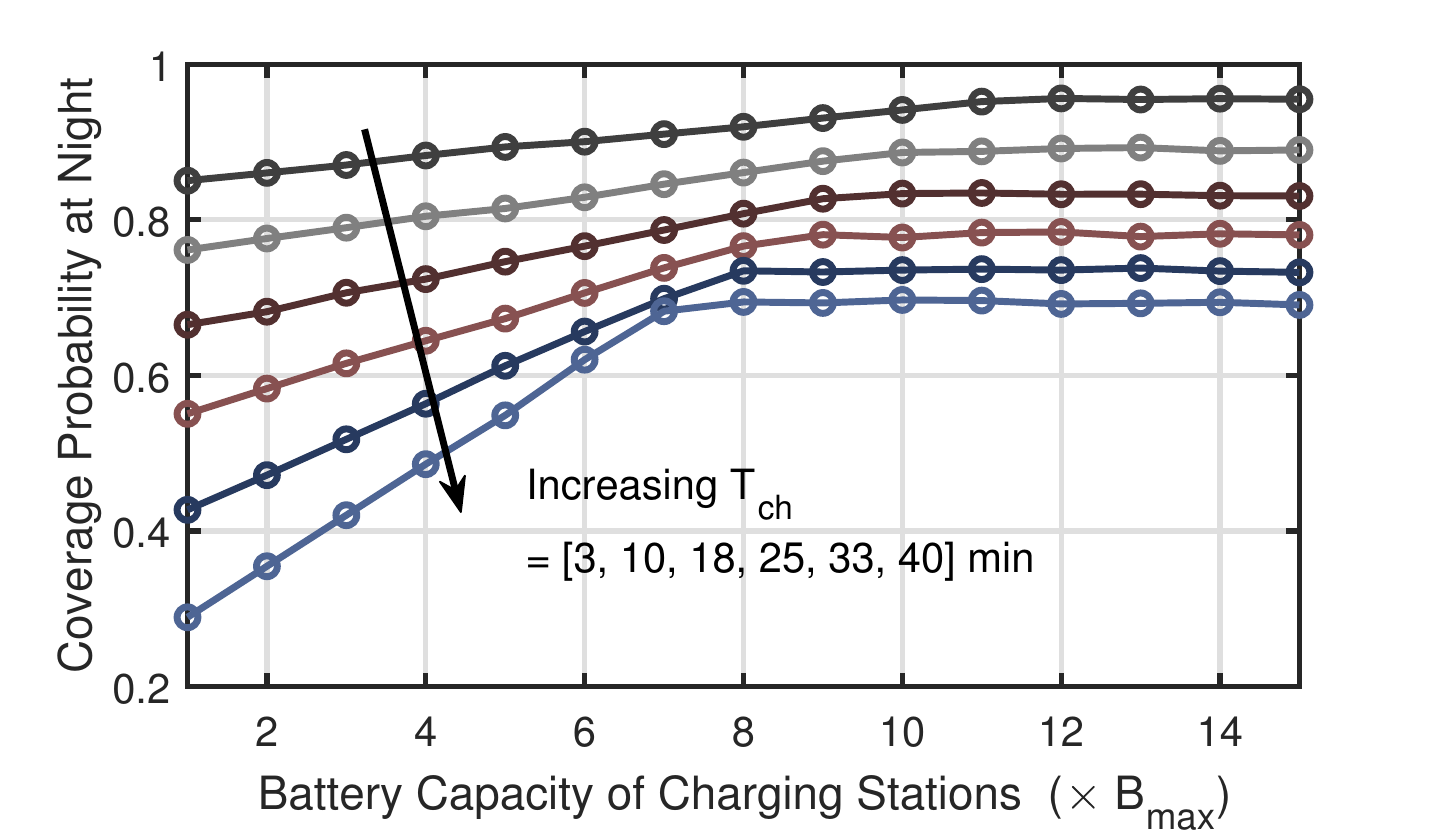}
	\caption{Coverage probabilities versus charging time under different capacities of the solar powered charging station.}
	\label{fig_cov_night}
\end{figure}

The result in Fig. \ref{fig_cov_night} reveals the influence of uncertainty of RE and limited battery capacity of charging station on the system performance.
%the sunlight radiation intensity and battery capacity of RE charging station.  
The coverage probability increases dramatically with the increase of RE charging stations' battery capacity, especially when the charging time is long, which shows the importance of deploying RE charging station, battery size and the availability of RE sources. To overcome the challenge during night-time and days with little sunlight, on the one hand, backup system and RE charging stations with larger battery capacity can be considered. On the other hand, optimizing the orientation of solar panels by using tracking system \cite{mouli2016system} to maximize the harvest energy is also a potential solution. However, such tracking system requires a trade-off between higher harvested energy and system complexity.

\subsection{Dynamic Movement of Rural Users}
\begin{figure}[ht]
	\centering
	\includegraphics[width=1\columnwidth]{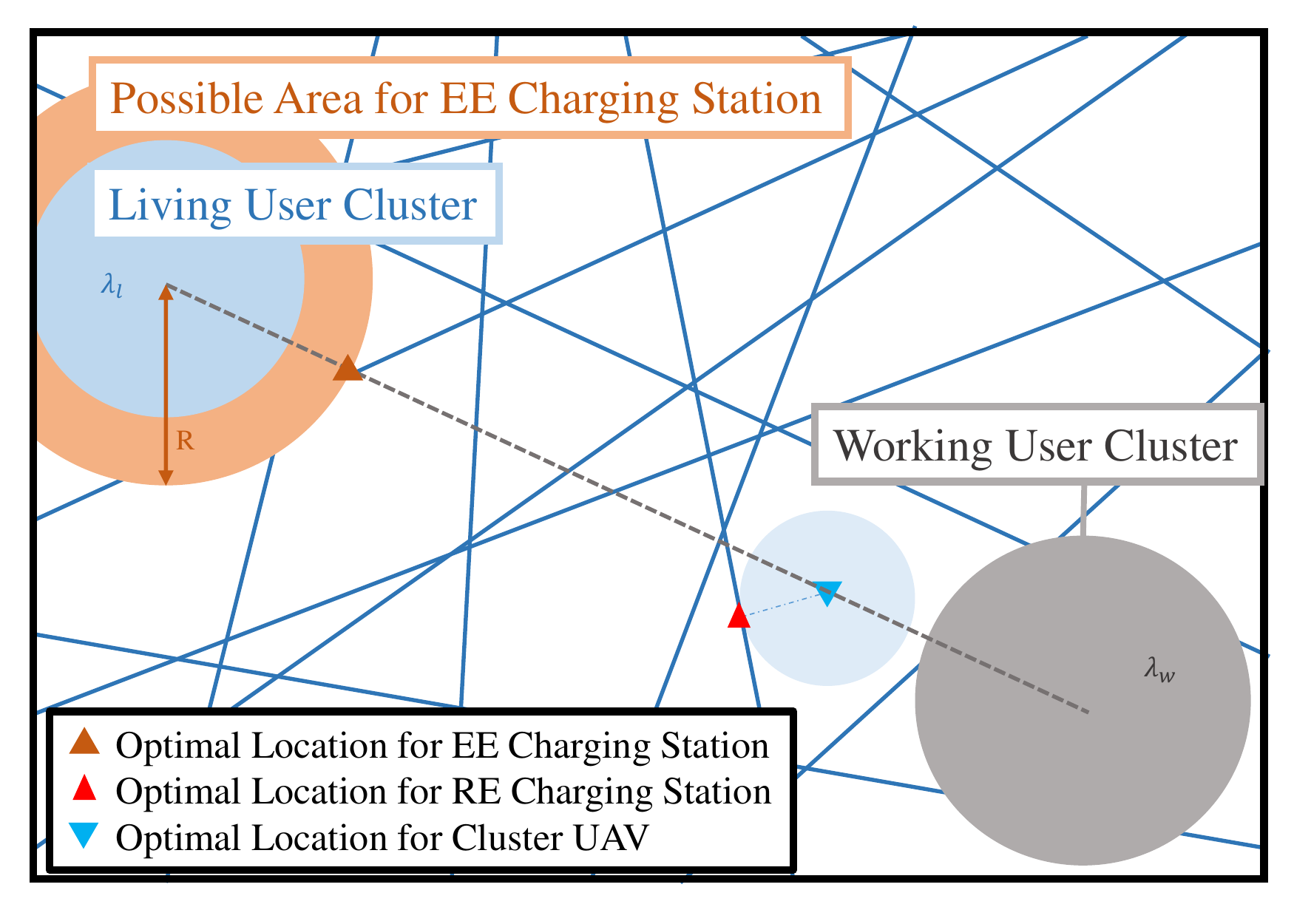}
	\caption{Illustration of system setup for Fig. \ref{fig_cov_distance}. The third scenario: a more realistic case which captures the dynamic movement of users.}
	\label{system_fig2}
\end{figure}
To capture the system performance in a more realistic model, we consider the medium-size user cluster in the first and second scenarios becomes the cluster pair  composed of two user clusters, one for living and another for working.  For example, some residents live in a village and work in a nearby farm. By using this model we can capture the dynamic movement of rural users. In this scenario, we consider one UAV for each medium-size user cluster pair. These two clusters are located at a fixed distance $d$ in a random orientation, with densities $\lambda_l$ and $\lambda_w$, respectively. Besides, the possible area of deploying EE charging station around the living cluster is constrained, and let $ R$ denotes the radius of the possible area. To be more realistic, we assume that the RE charging station can only be deployed on roads, of which the locations are modeled as a Poisson Line Process (PLP) \cite{dhillon2020poisson}, when it is outside of living clusters. The reason is that when deploying charging stations in the living cluster, it can be deployed on the top of the buildings if necessary, hence, it can locate almost everywhere in the living cluster. However, when it comes to farms and fields, there is no high dense building, so the possible area of deploying RE charging stations restricts to the road. In this scenario, we consider that the cluster UAV is deployed at the optimal location that maximizes the coverage probability of the medium-size user cluster pair, and the EE and RE charging stations are deployed accordingly, at their optimal locations that minimize the traveling distances of the cluster UAVs. For instance, if $\lambda_w$ is much higher than $\lambda_l$, the cluster UAV is much closer to the working cluster. In this case, the EE charging station is deployed at the edge of the living cluster, and the RE charging station is deployed at the nearest point on the road from the cluster UAV, as shown in Fig. \ref{system_fig2}.

\begin{figure}[ht]
	\centering
	\includegraphics[width=1\columnwidth]{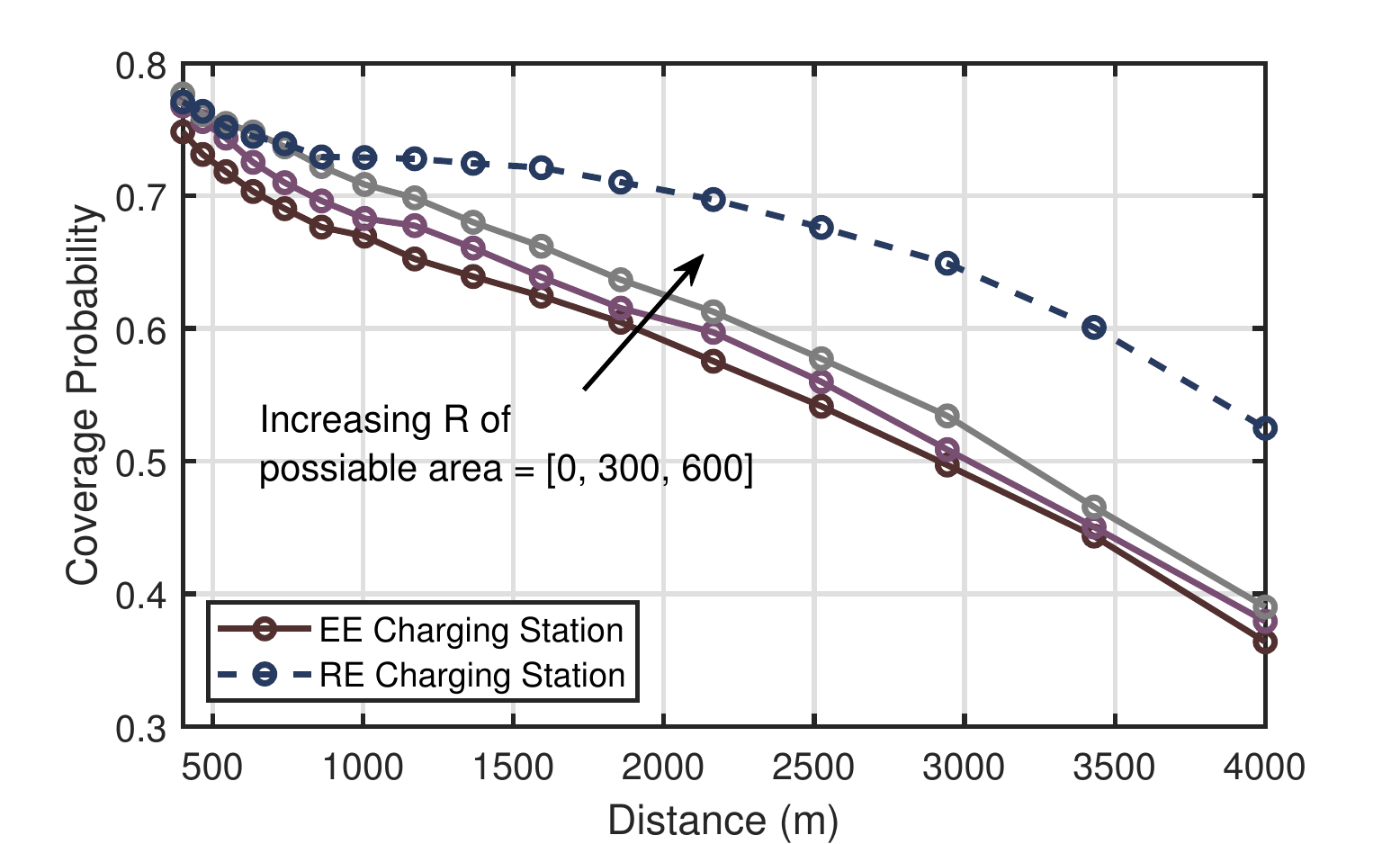}
	\caption{Coverage probabilities under different distances between working and living user cluster.}
	\label{fig_cov_distance}
\end{figure}
Fig. \ref{fig_cov_distance} shows a better system performance achieved by the deployment of RE charging stations when the distance of two user clusters is large and the performance of increasing the radius of possible area does not improve too much. The reason is the same as the first scenario: the deployment of RE charging stations highly reduce the traveling distance and energy consumed during traveling. Compared with energy from wired electricity, off-grid RE systems are more flexible, low maintenance cost and eco-friendly. Especially in rural areas, the possible areas of deploying RE charging stations are much wider than EE charging stations, and lower dense of high buildings makes it possible for the solar panel to harvest more energy. 

However, to achieve this performance, a large-scale deployment of RE charging stations is required because of the spatial dynamics of users and sparse distribution (e.g., the optimal locations of RE charging stations change with the change of locations of working clusters).
Despite of falling cost of solar panels \cite{mouli2016system}, lack of financing mechanisms coupled with a high start-up cost restrict the development of RE system in rural areas.  This problem can only be solved by paying more attentions on the development of rural areas and government support.

\section{Future Works and Challenges}
The work presented in this article captures the initial setup of designing RE charging stations and UAVs in enhancing the coverage in rural areas and opens up lots of issues that need to be addressed in the future research.

\subsection{UAV-enabled Networks}
While we show that combining UAVs and RE charging stations in the wireless network greatly improves system performance, it does not suit all situations. For example, low-altitude UAVs are greatly affected by climate change. As mentioned, rural areas have a high probability of facing extreme weathers, such as hurricanes, sand storms, and thunderstorms. Such harsh conditions can easily interrupt the service UAVs offer or even destroy UAVs. In addition, legislation of UAVs is still catching up. Since UAVs' wide use is relatively new, some uncertainties and conflicts about the airspace property rights and radio frequency interference should be clarified. Moreover, safety and privacy while dealing with UAVs are also primary and common concerns. UAVs can collect data and images while flying at a relatively high altitude without drawing attention. Therefore it leads to the problem of invasion of privacy and aerial surveillance, and UAVs systems may crash due to the latency made by remote controls or man-made mistakes.
 
\subsection{RE-enabled Networks}
In addition to the potential concerns derived by UAVs' networks, RE charging stations cannot suit all the scenarios, either. As mentioned, the high initial cost of installation restricts the deployment of RE-powered charging stations. Besides, the geographical limitations can impose extra set of challenges. Some renewable energy resources are simply not available in some remote areas. For example, the availability of sunlight in the Middle East is significantly more than in Central and Eastern Europe, meaning that different regions need different systems and  maybe combing more than one type of energy generation is more efficient and reliable in some areas. Moreover, such energy is generated by natural resources that are uncontrollable by humans. Wind energy depends on wind speed: if it is too low, the turbine does not turn, however, too much wind can damage the generator. These uncertainties make the integration more complex. Developing suitable size energy storage mechanisms is essential. How much energy should be stored such that the user can have a stable power supply and how to store the power in a cost-effective, reliable way now become the problems that need to be solved before the widespread of the RE-powered charging stations.

\subsection{Open Problems}
In the field of UAV-aided rural area wireless communications, future studies can be based on the optimal 3D locations of UAVs that maximize the coverage probability and trajectory optimization, which captures the dynamic and large-scale movement of rural users. While considering the locations of UAVs, the location of charging stations should also be investigated since UAVs cannot work without charging. Therefore, where to deploy and which kinds of charging stations should be deployed are promising directions for future work. Besides, different connectivity strategies based on different regions can be investigated. For instance, can we combine UAVs with HAPs or satellites, if so, what is the best UAV deployment strategy. Furthermore, a more realistic spatial distribution model of rural areas can be provided, such as modeling the dynamic movement of rural users and user clusters. 

There is no single solution that addresses challenges in all the rural areas. However, building on the current technologies and strategies while considering the particularities of each region and achieving ubiquitous connectivity is the most suitable solution to follow.

\section{Conclusion}
In this article, we investigated three practical scenarios of deploying RE charging stations in rural areas and analyzed its strengths and weaknesses numerically. Deploying RE charging stations do enhance the system performance dramatically. However, this system is easily influenced by weather, daytime duration (in case of solar-powered) and limited battery capacity of RE charging stations. Finally, we have enlisted a set of challenges and open problems that still need to be resolved before considering large scale deployment of RE-powered charging stations for drone-enabled systems in rural areas.

\bibliographystyle{IEEEtran}
\bibliography{Report4}
\vskip -2\baselineskip plus -1fil
\begin{IEEEbiographynophoto} 
	{Yujie Qin} is currently a MS/PhD student in the communication theory lab at King Abdullah University of Science and Technology (KAUST). 
	She received her B.Sc. degree from University of Science and Technology of China (UESTC) in 2020.
	Her current research interests include stochastic geometry.
\end{IEEEbiographynophoto}
\vskip -2\baselineskip plus -1fil
\begin{IEEEbiographynophoto} 
	{Mustafa A. Kishk}
	[S'16, M'18] is a postdoctoral research fellow in the communication theory lab at King Abdullah University of Science and Technology (KAUST). 
	He received his B.Sc. and M.Sc. degree from Cairo University in 2013 and 2015, respectively, and his Ph.D. degree from Virginia Tech in 2018. 
	His current research interests include stochastic geometry, energy harvesting wireless networks, UAV-enabled communication systems, and 
	satellite communications.
\end{IEEEbiographynophoto}
\vskip -2\baselineskip plus -1fil
\begin{IEEEbiographynophoto} 
	{Mohamed-Slim Alouini}
	[S'94-M'98-SM'03-F'09]  was born in Tunis, Tunisia. He received the Ph.D. degree in Electrical Engineering
	from the California Institute of Technology (Caltech), Pasadena,
	CA, USA, in 1998. He served as a faculty member in the University of Minnesota,
	Minneapolis, MN, USA, then in the Texas A$\&$M University at Qatar,
	Education City, Doha, Qatar before joining King Abdullah University of
	Science and Technology (KAUST), Thuwal, Makkah Province, Saudi
	Arabia as a Professor of Electrical Engineering in 2009. His current
	research interests include the modeling, design, and
	performance analysis of wireless communication systems.
\end{IEEEbiographynophoto}
\end{document}